\documentclass[aos,nameyear,dvips]{arximspdf}
\usepackage{dcolumn}
\usepackage{graphicx}

\doi{10.1214/10-AOS805}
\volume{38}
\issue{6}
\pubyear{2010}
\firstpage{3217}
\lastpage{3244}

\makeatletter

\newcolumntype{d}[1]{D{.}{.}{#1}}

\newtheorem{theorem}{Theorem}
\newtheorem{lemma}{Lemma}

\makeatother

\begin{document}
\begin{frontmatter}

\title{Gamma-based clustering via ordered means with application to
gene-expression analysis\thanksref{T1}}
\thankstext{T1}{Supported in part by NIH Grants R01 ES017400 and T32 GM074904.}
\runtitle{Gamma ranking}

\begin{aug}
\author{\fnms{Michael A.} \snm{Newton}\corref{}\ead[label=e1]{newton@stat.wisc.edu}} and
\author{\fnms{Lisa M.} \snm{Chung}\ead[label=e2]{lchung@stat.wisc.edu}}
\runauthor{M. A. Newton and L. M. Chung}
\affiliation{University of Wisconsin, Madison}
\address{Department of Statistics\\
University of Wisconsin, Madison\\
1300 University Ave\\
Madison, Wisconsin 53706-1532\\
USA\\
\printead{e1}\\
\phantom{E-mail: }\printead*{e2}} 
\end{aug}

\received{\smonth{7} \syear{2009}}
\revised{\smonth{12} \syear{2009}}

%
\begin{abstract}
Discrete mixture models provide a well-known basis for effective
clustering algorithms, although technical challenges have limited their
scope. In the context of gene-expression data analysis, a~model is
presented that mixes over a finite catalog of structures, each one
representing equality and inequality constraints among latent expected
values. Computations depend on the probability that independent
gamma-distributed variables attain each of their possible orderings.
Each ordering event is equivalent to an event in independent
negative-binomial random variables, and this finding guides a
dynamic-programming calculation. The structuring of mixture-model
components according to constraints among latent means leads to strict
concavity of the mixture log likelihood. In addition to its beneficial
numerical properties, the clustering method shows promising results in
an empirical study.
\end{abstract}

%
\begin{keyword}[class=AMS]
\kwd[Primary ]{60E15}
\kwd{62F99}
\kwd[; secondary ]{62P10}.
\end{keyword}
\begin{keyword}
\kwd{Gamma ranking}
\kwd{mixture model}
\kwd{next generation sequencing}
\kwd{Poisson embedding}
\kwd{rank probability}.
\end{keyword}

\end{frontmatter}

\section{Introduction}

A common problem in statistical genomics is how to
organize expression
data from genes that have been determined to exhibit differential
expression relative to various cellular states. Cells
in a time-course experiment may exhibit such genes, as may cells
in any sort of designed experiment or observational study where expression
alterations are being examined [e.g., \citet{parm03}, \citet{speed}].
In the event that the error-rate-controlled
list of significantly altered genes is small, the post-processing problem
amounts to inspecting observed patterns of expression, investigating
what is known about the relatively few genes identified, and planning follow-up
experiments as necessary. However, it is all too common that hundreds or
even thousands of genes are detected as significantly altered in their
expression pattern relative to the cellular states. Post-processing
these nonnull genes presents a substantial statistical
problem. Difficulties are compounded in the multi-group
setting because a
gene can be nonnull in many different ways [\citet{jensen}].

Ever since \citet{eisen}, clustering methods have been
used to organize expression data. Information about a
gene's biological function may be conveyed by the other genes sharing its
pattern of expression. \citet{clust06} provides
a recent
perspective. Clustering methods are often applied in order to partition
nonnull genes which have been
identified in differential expression analysis
[e.g., \citet{camp06}, \citet{gras08}].
Popular approaches
are informative but not completely satisfactory.
There are idiosyncratic problems, like
how to select the number of clusters, but there is also the subtle
issue that
the clusters identified by most algorithms are anonymous:
each cluster is defined only by similarity of its contents
rather than by some external pattern that its genes may be approximating.
Anonymity may contribute to
technical problems, such as
that the objective function being minimized is not
convex, and that realized clusters have a more narrow size distribution
than is warranted by the biological system.

Model-based clustering
treats data as arising from a mixture of component distributions, and then
forms clusters by assigning each data point to its most probable
component [e.g., \citet{tsm85},
\citet{mclabasf88}]. For example,
the \texttt{mclust} procedure is based
on mixtures of Gaussian components [\citet{fralraft02}];
the popular \texttt{K-means} algorithm
is implicitly so based [\citet{htf01}, page 463].
There is considerable flexibility in model-based clustering, though
technical challenges have also affected its development: the likelihood
function is often multi-modal; identifiability
can be difficult to establish
[e.g., \citet{redn84}, \citet{holzgnei06}];
and even where constraints may create identifiability, there can be a problem
of label-switching during Bayesian inference [\citet{step00}].
Some sophisticated model-based clustering methods have been developed for
gene expression [e.g., \citet{med1}]. Beyond empirical
studies, it is difficult to determine properties of such approaches,
and their
reliance on Monte Carlo computation is somewhat limiting.

Here a model-based clustering method is developed
that aims to support multi-group gene-expression analysis and
possibly other applications.
The method, called gamma ranking,
places genes in a cluster if their
expression patterns commonly approximate one element from a finite
catalog of possible structures, in contrast to anonymous methods
(Section~\ref{sec2}).
Under certain conditions, the component distributions are linearly independent
functions---each one associated with a structure in the finite
catalog---and this confers favorable computational characteristics to the gamma-ranking
procedure (Sections~\ref{sec4},~\ref{sec5}).
The cataloged structures record patterns of equality and
inequality among latent expected values.
Where normal-theory specifications seem to be intractable, a~gamma-based
mixture model produces closed formulas for all necessary component densities,
thanks to an embedding of the relevant gamma-distributed variables in a
set of Poisson processes (Section~\ref{sec3}). The formulation also extends
to Poisson-distributed responses that are characteristic of gene expression
measured by next-generation sequencing (Section~\ref{sec6}).

\section{Mixture of structured components}\label{sec2}

The data considered has a relatively simple layout. Each gene $g$
from a possibly large number is associated with a vector
$x_g = (x_{g,1}, x_{g,2}, \ldots, x_{g,n})$ holding measurements
of gene expression from $n$ distinct biological samples. The $n$
samples are distributed among $1 < p \leq n$ different groups, which
represent possibly
different transcriptional states of the cells under study.
The groups may represent cells exposed to $p$ different chemical
treatments, cells at $p$ different developmental stages, or cells
at $p$ different points along a time course, for example.
The layout of samples $\{1, 2, \ldots, n \}$ is recorded in
a vector, say $l = (l_1, l_2, \ldots, l_n )$, with $l_i = j$ indicating
that sample $i$ comes from group $j$. This is fixed by design and
known to the analyst; to simplify the development we suppress $l$ from
the notation below except where clarification is warranted.

Each expression
measurement $x_{g,i}$ is treated
as a positive, continuous variable representing
a fluorescence intensity from a microarray, after preprocessing has
adjusted for various systematic effects not related to the
groupings of interest.
Recent technological advances
allow expression to be measured instead as an explicit abundance count.
The mixture model developed below
adapts readily to this case (Section~\ref{sec6}).

Gamma ranking entails clustering genes according to the fit
of a specific model of gene-level data.
The joint
probability density for a data vector~$x_g$, denoted $p(x_g)$,
is treated as a finite mixture over a catalog of discrete
structures $\eta$, each of which determines
ordering constraints among latent expected values.
More specifically,
%
%
\begin{equation}
\label{eq:mix}
p(x_g) = \sum_{\eta} p(x_g|\eta) \pi_\eta,
\end{equation}
where
$\pi_\eta$ is a mixing proportion and the component density $p(x_g|\eta)$
is determined through modeling.

Each $\eta$ is a partition of group labels $\{1,2,\ldots,p \}$,
containing $K_\eta$ subsets, that also carries an ordering of these subsets.
For example,
three structures cover the two-group comparison, denoted
$\{ (1)(2), (12), (2)(1) \}$.
The notation conveys both
the partition of group means and the ordering of subsets within the
partition. For instance, in $\eta= (2)(1)$
the expected expression level in group $2$ is less than
that of group $1$; while $\eta= (12)$ indicates that both groups
share a common latent mean. With $p=3$ groups,
there are 13 structures
\begin{eqnarray*}
&& (123), (12)(3), (3)(12), (13)(2), (2)(13), (1)(23), (23)(1), \\
&& (1)(2)(3), (2)(1)(3), (1)(3)(2), (2)(3)(1), (3)(1)(2), (3)(2)(1),
\end{eqnarray*}
and the number grows rapidly with the number of groups (Table~\ref{tab0}).
A~way to think about $H_{\mathrm{ord}} = \{ \eta\}$, the catalog
of these ordered structures on
$p$ groups, is to imagine $p$ real values $y=(y_1, y_2, \ldots, y_p)$ and
the possible vectors you would get by ranking $y$. Of course there
are $p!$ rankings if ties are not permitted, but generally there are
far more rankings, and $H_{\mathrm{ord}}$ is in 1--1 correspondence
with the set of rankings of $p$ numbers, allowing ties.

%
\begin{table}
\tablewidth=220pt
\caption{The number of ordered structures, Bell$+$, as a function of the
number of groups, $p$. This is $\sum_{k=1}^p (k!) S(p,k)$, where
$S(p,k)$ are
Stirling numbers of the second kind. The Bell number of partitions of
$1,\ldots,p$ is included for comparison}\label{tab0}
\begin{tabular*}{\tablewidth}{@{\extracolsep{\fill}}lcc@{}}
\hline
$\bolds p$ & \multicolumn{1}{c}{\textbf{Bell}$\bolds{+}$} &
\multicolumn{1}{c@{}}{\textbf{Bell}} \\
\hline
2 & \phantom{000}3 & \phantom{00}2 \\
3 & \phantom{00}13 & \phantom{00}5 \\
4 & \phantom{00}75 & \phantom{0}15 \\
5 & \phantom{0}541 & \phantom{0}52 \\
6 & 4683 & 203 \\
\hline
\end{tabular*}
\end{table}

An ordered structure $\eta$ also dictates an association between sample
labels $i \in\{1,2, \ldots, n\}$ and levels of the latent expected
values. The null structure $\eta=(12\cdots p)$,
for example, entails equal mean expression
across all $p$ groups; all observations are associated with a single
mean value (and we write $K_\eta= 1$).
More generally, there are $K_\eta>1$ distinct mean values,
$\mu_1 < \mu_2 < \cdots< \mu_{K_\eta}$, say. Without loss of generality,
we index the means by rank order. The association maps
each $i \in\{1, 2, \ldots, n \}$ to some $\mu_k$; it amounts
to a partition of the samples together with an ordering
of the subsets within the partition matching the order of
the latent means. We express this association with
disjoint subsets $\sigma(\eta,k)$, $k=1,2,\ldots, K_\eta$, and have
$k$ follow the order of the expected values. For
example, suppose that samples $\{1,2, \ldots, 6\}$ constitute two replicate
samples in each of $p=3$ groups, and
$\eta=(13)(2)$ is considered to relate the group-specific expected values
(i.e., the gene is upregulated in group 2, and not differentially
expressed between groups~1 and 3).
Then $K_\eta=2$,
$\sigma(\eta,1)=\{1,2,5,6\}$ and $\sigma(\eta,2)=\{3,4\}$.
Subset $\sigma(\eta,k)$
includes $n_k$ samples and induces gene-level statistics such as
\[
s_{g,k} = \sum_{ i \in\sigma(\eta,k) } x_{g,i}
\quad\mbox{and}\quad
t_{g,k} = \prod_{i \in\sigma(\eta,k) } x_{g,i} .
\]

The structure/partition notation
is convenient in multi-group mixture modeling. For clarification,
let us refer back to the layout notation and
take the replicates $r_j = \{i\dvtx l_i = j \}$, which equal those
samples in group $j$.
Consider a gene that is completely differentially expressed relative to
the $p$ groups; that is, it assumes one of the $p!$ structures
$\eta$ in which $K_\eta= p$. It follows that each set $r_j$ equals
exactly one of the subsets $\sigma(\eta,k)$. [It would be $\sigma(\eta,1)$
if $r_j$ had the lowest mean expression level, e.g.]
In the absence of complete differential expression, multiple groups
share expected values. Generally, therefore, each subset $\sigma(\eta
,k)$ is
a union of various replicate sets~$r_j$.
The language also conveys the assumption
that replicates $i_1$ and $i_2$ in the same set $r_j$ necessarily share
expected value, regardless of the structure $\eta$.
In calculating probabilities,
the sets $\sigma(\eta,k)$ of equi-mean samples are more important
than the replicate sets $r_j$.

From the mixture model~(\ref{eq:mix}), posterior structure
probabilities are
$p(\eta|x_g) = p(x_g|\eta) \pi_\eta/p(x_g)$ and these determine
gene clusters by Bayes's rule assignment. Alternatively,
the cluster contents can be regulated by a threshold parameter $c$, and
%
%
\begin{equation}
\label{eq:reg}
\operatorname{cluster}(\eta) = \{ g\dvtx p(\eta|x_g) \geq c \},
\end{equation}
though some genes may go unassigned in this formulation.
In any case, each cluster holds genes with empirical
characteristics matching some discrete mean-ordering structure.


The latent expected values are constrained by
$\eta$ to the order $\mu_1 < \mu_2 <\cdots< \mu_{K_\eta}$.
Propeling our calculations is the ability to integrate these
ordered means
(i.e., marginalize them) in a model involving gamma distributions
on some
transformation of the $\mu_k$'s. Recall that a gamma distribution
with shape $a>0$ and rate $\lambda>0$, denoted Gamma$(a,\lambda)$,
has probability density
\[
p(z) = \frac{ \lambda^{a} z^{a-1} \exp\{ -z \lambda\} }{\Gamma(a)
},\qquad
z>0.
\]
We assume that inverse means $\psi_k = 1/\mu_k$ have joint density
%
%
\begin{eqnarray}
\label{eq:comp1}
p_\eta(\psi_1, \ldots, \psi_{K_\eta} )
&=& K_\eta! \Biggl[
\prod_{k=1}^{K_\eta} \frac{ (\alpha_0 \nu_0)^{\alpha_0} \psi_k^{\alpha
_0 -1}
\exp\{ -\alpha_0 \nu_0 \psi_k \} }{
\Gamma( \alpha_0 ) } \Biggr] \nonumber\\[-8pt]\\[-8pt]
& &\hspace*{0pt}{} \times1 [ \psi_1 > \psi_2 >\cdots> \psi_{K_\eta}
],\nonumber
\end{eqnarray}
which reflects independent and identically distributed
Gamma$(\alpha_0, \alpha_0 \nu_0)$ components, conditioned to
one ordering. This parameterization gives $\nu_0$ an interpretation
as a centering parameter; on the null structure
having a single latent mean $\mu_1$, $1/\nu_0 = E ( 1/\mu_1 )$.

To complete the hierarchical specification, we
assume a gamma observation model
%
%
\begin{eqnarray}
\label{eq:comp2}
p(x_g|\psi_1, \ldots, \psi_{K_\eta},\eta) &=&
\prod_{k=1}^{K_\eta} \prod_{i \in\sigma(\eta,k) }
\frac{ ( \alpha\psi_k )^{\alpha} x_{g,i}^{\alpha-1} \exp\{
- x_{g,i} \psi_k \alpha\} } { \Gamma(\alpha) } \nonumber\\[-8pt]\\[-8pt]
&=& \prod_{k=1}^{K_\eta}
\frac{ ( \alpha\psi_k)^{\alpha n_{k} } t_{g,k}^{\alpha-1} \exp\{
- s_{g,k} \psi_k \alpha\} } { (\Gamma(\alpha))^{n_k} }.\nonumber
\end{eqnarray}
Equivalently, with sample $i \in\sigma(\eta, k)$, measurement
$x_{g,i}$ is
distributed as Gamma$(\alpha, \alpha\psi_k)$, all conditionally on
the latent values and $\eta$, and independently across samples.
The gamma observation component is often
supported empirically; there is theoretical
support from stochastic models of population abundance [\citet{dennpati84},
\citet{remppawl08}] and there are practical
considerations that a gamma-based model may be the only one for
continuous data in which
ordering calculations are feasible.

The structured component $p(x_g|\eta)$ in~(\ref{eq:mix})
arises by integrating~(\ref{eq:comp2})
against the continuous mixing distribution~(\ref{eq:comp1}).
Specifically,
\[
p(x_g|\eta) = \int p(x_g| \psi_1, \ldots, \psi_{K_\eta}, \eta)
p_\eta(\psi_1, \ldots, \psi_{K_\eta} ) \,d\psi_1 \cdots
d\psi_{K_\eta} .
\]
Moving allowable factors from the integral
\begin{eqnarray*}
p(x_g|\eta) &=& \frac{K_\eta! (\alpha_0 \nu_0)^{K_\eta\alpha_0} \alpha
^{ \alpha n}
}{ \Gamma(\alpha_0)^{ K_\eta} \Gamma(\alpha)^n }
\Biggl( \prod_{k=1}^{K_\eta} J_k t_{g,k}^{\alpha- 1} \Biggr) \\
& &{} \times
\int_E \prod_{k=1}^{K_\eta} \frac{ \psi_k^{ \alpha_0 + \alpha n_k -1 }
\exp\{ -\psi_k ( \alpha_0 \nu_0 + \alpha s_{g,k} )\} }
{J_k} \,d\psi_1 \cdots d{\psi_{K_\eta}},
\end{eqnarray*}
where the integral is over the set $E$ of decreasing $\psi_k$'s,
and where $J_k$
represents any cluster-specific quantity which does not depend on $\psi_k$.
Choosing
\[
J_k = \frac{ \Gamma( \alpha_0 + \alpha n_k ) }{ (\alpha_0 \nu_0
+ \alpha s_{g,k})^{ \alpha_0 + \alpha n_k } }
\]
provides just the right normalization, because then
the integrand becomes the joint
density of independent gamma-distributed variables, with the $k$th variable
having shape $a_k = \alpha_0 + \alpha n_k$ and rate $\lambda_k =
\alpha_0 \nu_0 + \alpha s_{g,k}$. The integral itself, denoted
$P_{\mathrm{ord}}(\eta)$, is
the probability that independent gamma-distributed variables assume a
certain order.
The preceding factor can be arranged as products
of the product statistics $t_{g,k}$ multiplied by factors involving the
sum statistics $s_{g,k}$. After a bit of simplification,
the following result is established.
\begin{theorem}\label{theo1}
In the model defined above, the component density $p(x_g|\eta)$ equals
%
%
\begin{equation}
\label{eq:gg}
c_\eta \Biggl( \prod_{i=1}^n x_{g,i}^{\alpha- 1} \Biggr)
\underbrace{ \prod_{k=1}^{K_\eta} \biggl( s_{g,k} + \frac{\alpha_0 \nu
_0}{\alpha}
\biggr)^{ -a_k } }_{ \mathrm{center} (\eta) }
\underbrace{ P (Z_1 > Z_2 > \cdots> Z_{K_\eta} ) }_{ P_{\mathrm{ord}}(
\eta) },
\end{equation}
where the $Z_k$'s are mutually
independent gamma-distributed random variables with
shapes $a_k = \alpha_0 + \alpha n_k$ and rates
$\lambda_k = \alpha_0 \nu_0 + \alpha s_{g,k}$, and where the
normalizing constant is
\[
c_\eta= \frac{K_\eta!}{ [ \Gamma(\alpha) ]^n
[ \Gamma(\alpha_0) ]^{ K_\eta} }
\biggl( \frac{ \alpha_0 \nu_0 }{ \alpha} \biggr)^{\alpha_0 K_\eta}
\prod_{k=1}^{K_\eta} \Gamma( a_k ) .
\]
In~(\ref{eq:gg}), $P_{\mathrm{ord}}( \eta) = 1$ for the null case involving
$K_\eta= 1 $.
\end{theorem}

The null structure $\eta=(12\cdots p)$ entails equal mean expression for
all samples; there is a single partition element, and $K_\eta= 1$.
In this case, the
distribution in~(\ref{eq:gg})
is exchangeable and equals a multivariate compound gamma
[\citet{hutch81}].
The positive parameters $\alpha$ and
$\alpha_0$ regulate within-group and among-group variation,
and $\nu_0$ is a scale parameter.
Inspection also confirms that if the
random $X=(X_1, \ldots, X_n)$ has density $p(x|\eta)$ in~(\ref{eq:gg}), and
if $b>0$, then $Y=(bX_1, \ldots, bX_n)$ has a density of the same type,
with shape parameters $\alpha_0$ and $\alpha$ unchanged, but with scale
parameter $b \nu_0$.

Special cases of the density~(\ref{eq:gg}) have been reported:
\citet{newt04} presented the case $p=2$; \citet{jensen} presented the case $p=4$. See also
\citet{yuankend06a}.
Evidently an algorithm to compute $P_{\mathrm{ord}}(\eta)$
is required in order to evaluate the component mixing
densities. Beyond the $p=2$ case, previous reports have
evaluated these gamma-rank probabilities by Monte Carlo.

%
\begin{figure}

\includegraphics{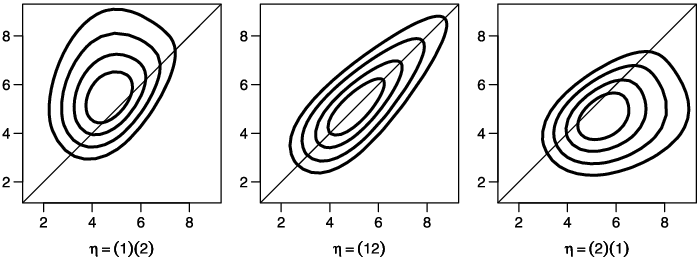}

\caption{Three structured components in $\mathbb{R}^2$.
Here $\alpha=10$, $\alpha_0=3$ and $\nu_0= 2^5$.
Contours cover 50\%, 80\%, 95\%
and 99\% probability. For convenience, each density is shown for
$\log_2$-transformed pairs.}\vspace*{-3pt}
\label{fig2}
\end{figure}

Figure~\ref{fig2} displays contours of the three structured components
when $n=2$ and $p=2$. Clearly the components
distribute mass quite differently from one another, and in a way that reflects
constraints encoded by $\eta$. The densities from
different structures $\eta$ have the
same support; the constraints restrict latent expected values rather
than observables.
In this way, the approach shares something with generalized linear modeling
wherein responses are modeled by generic
exponential family densities and covariate information constrains the
expected values [\citet{mccuneld89}].

\section{Gamma-rank probabilities}\label{sec3}

A statistical computing problem must be solved
in order to implement gamma ranking. Specifically, it is required to
calculate the probability $P(E)$ of the event
%
%
\begin{equation}
\label{eq:e}
E = \{ Z_1 > Z_2 > \cdots> Z_K \},
\end{equation}
where $\{ Z_k\dvtx k=1,2,\ldots, K \}$
are mutually independent gamma distributed random
variables with possibly different shapes
$a_1, a_2, \ldots, a_K$ and rates $\lambda_1, \lambda_2, \ldots,
\lambda_k$.
[Each $P_{\mathrm{ord}}(\eta)$ in Section~\ref{sec2} is an instance of $P(E)$.]
In the special case $K=2$, the event in
two gamma-distributed variables is equivalent to the $E' =
\{ B > \lambda_1/(\lambda_1 + \lambda_2) \}$,
where $B$ is a Beta$(a_1,a_2)$ distributed variable.
Thus, $P(E)=P(E')$ can be computed
by standard numerical approaches for the Beta distribution.
Although a similar representation
is possible for Dirichlet-distributed vectors when $K>2$, a~direct
numerical
approach is not clearly indicated.
In modeling permutation data, \citet{stern90}
presented a formula for $P(E)$ for any value $K$, but
assuming common shape
parameters $a_k = a$. \citet{sobel} calculated $P(E)$ for
$K < 5$ and assuming constant rates $\lambda_k = \lambda$, but to our
knowledge a general formula has not been developed. A~Monte Carlo approximation is certainly feasible, but
a fast and accurate
numerical approach would be preferable for computational efficiency:
target values may be small, and
$P(E)$ may need to be recomputed for many shape and rate settings.

There is an efficient numerical approach to computing $P(E)$ when
shapes $a_k$ are positive integers. The approach involves
embedding $ \{ Z_k \}$ in a collection
of independent Poisson processes $ \{ \mathbb{N}_k \}$,
where $k=1,2, \ldots, K$. Specifically,
let $\mathbb{N}_k$ denote a Poisson process on $(0, \infty)$ with rate
$\lambda_k$. So $\mathbb{N}_k(0,t] \sim\operatorname{Poisson}( t \lambda
_k )$,
for example. Of course, gaps between points in $\mathbb{N}_k$ are
independent and
exponentially distributed, and the gamma-distributed
$Z_k$ can be constructed by summing the first $a_k$ gaps
\[
Z_k = \min\{ t>0\dvtx\mathbb{N}_k(0,t] \geq a_k \}.
\]
Next, form\vspace*{1pt} processes $ \{ \mathbb{M}_k \}$ by
accumulating points in the originating processes:
$ \mathbb{M}_k = \sum_{j=1}^k \mathbb{N}_j $.
Marginally, $\mathbb{M}_k$ is a Poisson process with rate
$\Lambda_k = \sum_{j=1}^k \lambda_j$, but over $k$ the
processes are dependent owing to overlapping points.
To complete the construction, define
count random variables $M_1, M_2, \ldots,\break M_{K-1}$ by
%
%
\begin{equation}
\label{eq:def1}
M_k = \mathbb{M}_k ( 0, Z_{k+1} ].
\end{equation}
It is immediate that each $M_k$ has a marginal negative
binomial distribution: the gamma distributed
$Z_{k+1}$ is independent of $\mathbb{M}_k$;
conditioning on $Z_{k+1}$
in~(\ref{eq:def1}) gives a Poisson variable which mixes out to the
negative binomial [\citet{gy20}]. Specifically,
\[
M_k \sim\operatorname{NB} (\mathrm{shape} =a_{k+1}, \mathrm{scale}=
\Lambda_k/\lambda_{k+1} ),
\]
which corresponds to the probability mass function
%
%
\begin{equation}
\label{eq:nb}
p_k(m) = \frac{ \Gamma( m+a_{k+1} ) }{ \Gamma( a_{k+1} ) \Gamma(m+1) }
\biggl( \frac{\lambda_{k+1}}{\Lambda_{k+1}} \biggr)^{a_{k+1}}
\biggl( \frac{\Lambda_k}{\Lambda_{k+1} } \biggr)^{m}
\end{equation}
for integers $m \geq0$.
The next main finding is
the following.
\begin{theorem}\label{theo2}
With $E$ as in~(\ref{eq:e}), $M_k$ as in~(\ref{eq:def1})
and $p_k$ as in~(\ref{eq:nb}), $P(E)$ equals
%
%
\begin{equation}
\label{eq:f1}
\sum_{m_1=0}^{a_1 - 1} \sum_{m_2=0}^{m_1+a_2 -1 }
\cdots\sum_{m_{K-1} = 0 }^{m_{K-2} + a_{K-1} - 1 }
p_1(m_1) p_2(m_2) \cdots p_{K-1}(m_{K-1}).
\end{equation}
\end{theorem}

It does not seem to be obvious
that $E$ in~(\ref{eq:e}) is equivalent to an event in the $\{M_k \}$.
We also find it striking that the $M_k$ variables are independent
considering that they are constructed from highly dependent $\mathbb{M}_k$
processes. Proof of~(\ref{eq:f1}) and the related distribution theory
are presented in Appendix~\ref{appA}.

A redistribution of products and sums
allows a numerically efficient evaluation of~(\ref{eq:f1}), as in
the sum-product algorithm [e.g.,
\citet{fg}]. For instance, with $K=4$,
%
%
\begin{equation}
\label{eq:f2}
P(E) = \sum_{m_1=0}^{a_1 - 1 } p_1(m_1) \Biggl\{
\sum_{m_2 = 0}^{m_1+a_2-1} p_2(m_2) \Biggl[
\sum_{m_3 = 0}^{m_2 + a_3-1} p_3(m_3) \Biggr] \Biggr\}.
\end{equation}
Here, one would evaluate $P(E)$ by first constructing for each $m_2 \in
\{ 0, 1, \ldots,\break a_1+a_2-2 \}$ an inner sum $P( M_3 \leq m_2+a_3-1 )$.
This vector in $m_2$ values is used to process the second inner sum,
for each value $m_1 \in\{0,1, \ldots, a_1-1\}$. Indeed the computation
is completely analogous to the Baum--Welch backward recursion
[e.g., \citet{rabiner}], although,
interestingly, there seems to be no hidden Markov chain in the system. A~version of
the Viterbi algorithm identifies the maximal summand and thus provides an
approach to computing $\log P(E)$ in case $P(E)$ is very
small.

\section{Linear independence}\label{sec4}

The component densities~(\ref{eq:gg}) seem to have the useful property
of being
linearly independent functions on $\mathbb{R}^n$. Linear
independence of the component density functions is equivalent to
identifiability of
the mixture model [\citet{ys68}].
It is necessary for strict concavity of
the log-likelihood, but it is not routinely established. Establishing
identifiability also is a key step in determining sampling properties
of the maximum likelihood estimator.

Let $a=(a_\eta)$ denote a vector of real numbers indexed by structures
$\eta$.
Recall that the finite catalog of functions $\{ p(x|\eta) \}$
is linearly independent if
\[
T_a(x) = \sum_\eta a_\eta p(x|\eta) = 0 \qquad\mbox{for all $x$
implies $a_\eta=0$ for all $\eta$}.
\]
It is plausible that
this property holds generally, but we have been able to establish
a proof only in a special case.
\begin{theorem}\label{theo3}
In a balanced experiment where $m$ replicate samples are measured in
each of $p=2$ or $p=3$ groups, the component densities $p(x_g|\eta)$ in
(\ref{eq:gg}) are linearly independent functions on $\mathbb{R}^{mp}$.
\end{theorem}

A proof proceeds by finding
a multivariate polynomial $\phi(x) > 0$ such that $\phi(x) T_a(x) $
is itself a multivariate polynomial. A~close study of
the degrees and coefficients of this polynomial leads us to the result
(Appendix~\ref{appB}).
That such a $\phi(x)$ exists follows from~(\ref{eq:gg}): the center
is a rational function, and the factor
$P_{\mathrm{ord}}(\eta)$ is also rational, being
a linear combination of rational functions, as established in
(\ref{eq:f1}).

\section{Data analysis considerations}\label{sec5}

\subsection{Estimation}\label{sec51}

To deploy model~(\ref{eq:mix})--(\ref{eq:gg})
requires the estimation of parameters $\alpha$,
$\alpha_0$ and $\nu_0$, which are shared by the different components,
as well as mixing proportions $\pi= \{ \pi_\eta\}$,
which link the components together. Consider first the log likelihood
for $\pi$ alone (treating the shared parameters as known)
under independent and identically distributed sampling from~(\ref{eq:mix}):
%
%
\begin{equation}
\label{eq:loglik}
l(\pi) = \sum_{g=1}^G \log\biggl\{ \sum_{\eta} \pi_\eta p(x_g|\eta)
\biggr\},
\end{equation}
where $G$ is the number of genes providing data.
Maximum likelihood estimation of $\pi$ is buttressed
by the following finding.
\begin{theorem}\label{theo4}
Suppose that the component densities are linearly independent functions
in the mixture of structured components model.
If $G$ is sufficiently large, then
the log likelihood $l(\pi)$ in~(\ref{eq:loglik})
is strictly concave on
a convex domain, and thus admits a unique maximizer $\hat\pi=
\{ \hat\pi_\eta\} $. This property is almost sure in data sets.
\end{theorem}

The expectation--maximization (EM) algorithm naturally applies
to approximate~$\hat\pi$.
By strict concavity of $l(\pi)$, it is not necessary to rerun
EM from multiple starting points. The final estimate and resultant
clustering should be insensitive to starting position, as has been found
in numerical experiments. This is a convenient but unusual property in
the domain of mixture-based clustering [\citet{mclapeel00}, page 44].

In a small simulation experiment, we confirmed
that our implementation of the EM algorithm was able to recover mixture
proportions given sufficiently many draws from the marginal
distribution~(\ref{eq:mix})
(data not shown).

Full maximum likelihood for both the mixing proportions and shared
parameters is feasible via the EM algorithm, but this increases
computational costs. In the prototype implementation used here, we
fixed the shared parameters at estimates obtained from a simpler
mixture model, and then ran the EM algorithm to estimate the
mixing proportions. Specifically, we used the gamma--gamma method
in \texttt{EBarrays} (\href{http://www.bioconductor.org}{www.bioconductor.org}),
which corresponds to mixing as in~(\ref{eq:mix}) but
over the smaller set of unordered structures. Experiments
indicated that this approximation had a small
effect on the identified clusters (see Appendix~\ref{appD}).

Inference derived through the proposed parametric model is reliant to
some degree on the validity of the governing assumptions. Quantile--quantile
plots and plots relating sample coefficient-of-variation to sample mean
provide useful diagnostics for the gamma observation-component of the model.
The within-component model is restrictive in the sense that three
parameters are shared among all the components (i.e., structures).
This can be checked
by making comparisons of inferred clusters, but only large clusters
would deliver any power. Clusters reveal patterns in mean expression,
while the shared parameters have more to do with variability; if other
domains of statistics provide a guide, one expects that misspecifying the
variance may reduce some measure of
efficiency without disabling the entire procedure.
The ultimate issue is whether or not the clustering method usefully
represents any underlying biology. This is difficult to assess, though
we examine the issue in a limited way in the examples considered next.

%
\begin{figure}

\includegraphics{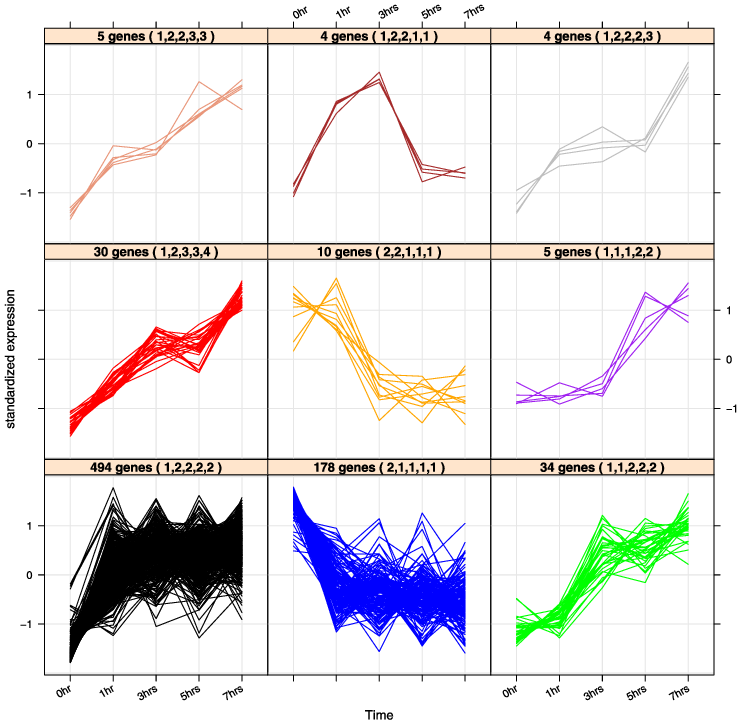}

\caption{Dominant patterns of differential expression in time course data
from Edwards et al. (\protect\citeyear{edwards}). Each panel summarizes data from one
cluster identified by gamma ranking (the nine largest clusters are shown). A~digital code signifies the inferred ordering of the latent expected
values (i.e., $\eta$, in an alternative notation). Each gene is a
single line trace; triplicate measurements were reduced by averaging and
then standardized for display; raw data went into the model fitting.
Results are based on 100 cycles of EM
to estimate mixing proportions followed by Bayes' rule assignment.}
\label{figure2}\vspace*{-3pt}
\end{figure}

\subsection{Example}\label{sec52}
\citet{edwards} studied the
transcriptional response of mouse heart tissue to oxidative stress.
Three biological replicate samples were measured using Affymetrix
oligonucleotide arrays at each of five time points (baseline and
one, three, five and seven hours after a stress treatment) for several
ages of mice. Considering the older mice
for illustration, we have
$p=5$ distinct groups, $n=15$ samples and 10,043 genes (i.e.,
probe sets, after pre-processing).
Gene-specific moderated F-testing [\citet{smyth04}]
produced a list of $G=786$ genes
that exhibited a significant temporal response to stress at the $10\%$
false discovery rate [by $q$-value; \citet{stotibs03}].
Gamma ranking involved fitting the mixture of
structured components,\vadjust{\goodbreak} which with $p=5$ mixes over $540$ distinct components.
(Since we worked with significantly altered genes, we did not include the
null component in which all means are equal; other aspects of model
fitting and diagnosis are provided in Appendix~\ref{appD}.)
From the catalog of $540$ possibilities, genes populated 23 clusters
by gamma ranking, though only four clusters contained 10 or more of the
$G=786$ stress-responding genes (Figure~\ref{figure2}).
Most expression changes occurred between baseline and
the first time point,
but 30 genes (red cluster)
showed significant up-regulation at all but one time point, for example.\vadjust{\goodbreak}

Gamma ranking gave different results than
\texttt{K-means} or \texttt{mclust}, which, respectively, found $20$
and $2$
clusters in Edwards' data.
Here \texttt{K} was chosen according to guidelines in \citet{htf01},
\texttt{mclust} used the Bayes information criteria over
the range from 1 to 50 clusters. Otherwise, both
methods used default settings in the respective R functions
(\href{http://www.r-project.org}{www.r-project.org}).
The adjusted Rand index [\citet{ha85}], which measures
dissimilarity of partitions,
was $0.09$ comparing gamma ranking and \texttt{K-means},
$0.16$ for gamma ranking and \texttt{mclust}, while for
\texttt{K-means} and \texttt{mclust} it was smaller, at $0.02$.

The biological significance of clusters identified by any algorithm
may be worth investigating. For example, the cluster of
30 increasing expressors includes 2 genes (Mgst1 \& Gsta4) from
among only 17 in the whole genome that are involved
in glutathione transferase activity.
Understanding the
increased activity of this molecular function will give a more complete
picture of the biology [e.g., \citet{gira}].
In isolation, it is difficult to see how such investigation is supportive
of a given clustering approach. The benefits become more apparent
when we look at many data sets and many functional categories.

\subsection{Empirical study}\label{sec53}
\begin{figure}[b]
\vspace*{-3pt}
\includegraphics{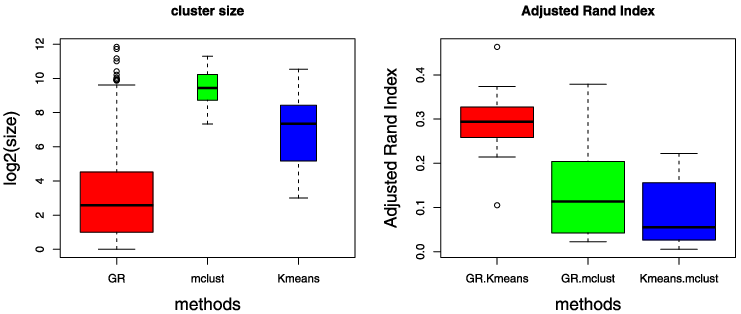}

\caption{Characteristics of clusters from an empirical study of 11 data sets.}
\label{figure3}
\end{figure}

Gamma ranking was applied to a series of 11 data sets obtained
from the Gene Expression Omnibus (GEO) repository [\citet{edgar}].
These were all the data sets satisfying a specific and relevant query
(Table~\ref{tab:dat}). They represent experiments
on different organisms and they exhibit a range of variation characteristics.
In each case, we applied
the moderated F-test and selected genes with $q$-value no larger than $5\%$.
Gamma ranking and, for comparison,
\texttt{mclust} and \texttt{K-means}, were applied in order to
cluster genes separately for each data set.
Basic facts about the identified clusters are reported in Table~\ref{tab:dat}.
Figure~\ref{figure3} shows that gamma ranking tends to produce smaller clusters than
\texttt{mclust} and \texttt{K-means}, although it also has a wider size
distribution;
and there was a relatively low level of
overlap among the three approaches.

%
\begin{table}
\caption{Summary of 11 data sets from the Gene Expression Omnibus (GEO).
GDS is the GEO data set accession number. These sets satisfied
the search query from August 2008
having subset variable type \textup{time} or \textup{development
stage} or \textup{age} and having a single factor with three to eight levels.
$p$~indicates the number of groups and $n$ is the number of samples.
$G$~indicates the number of genes deemed significantly altered by one-way
moderated F-test and 0.05 FDR (limma). The remaining columns show
how many clusters are found by gamma ranking with
100 EM iterations (GR), mclust (MC) and
K-means (KM)}\label{tab:dat}
\begin{tabular*}{\tablewidth}{@{\extracolsep{\fill}}d{4.0}cccd{2.0}d{5.0}d{3.0}cc@{}}
\hline
\textup{\textbf{GDS}} & \multicolumn{1}{c}{\textbf{Citation}} &
\multicolumn{1}{c}{\textbf{Organism}} & \multicolumn{1}{c}{$\bolds{p}$}
& \multicolumn{1}{c}{$\bolds{n}$} &
\multicolumn{1}{c}{$\bolds{G}$} & \multicolumn{1}{c}{\textbf{GR}}
& \multicolumn{1}{c}{\textbf{MC}} & \multicolumn{1}{c@{}}{\textbf{KM}}
\\
\hline
2323&Coser et al. &Homo Sapiens & 3&9& 1409& 11& 5& 13\\
1802&Tabuchi et al. &Mus musculus& 4&8& 3433& 49& 7& 10\\
2043&Tabuchi et al. &Mus musculus& 4&8& 3001& 51& 8& 18\\
2360&Ron et al. &Mus musculus& 4&9& 8714& 50& 8& 30\\
599&Vemula et al. &Rattus norvegicus&5&10& 673& 42& 2& 40\\
812&Zeng et al. &Mus musculus& 5&17& 10\mbox{,}982& 135& 7& 15\\
1937&Pilot et al. &Drosophila &5&15& 7733& 88& 8& 10\\
568&Welch et al. &Mus musculus& 6&18& 3737& 134& 4& 25\\
2431&Keller et al. &Homo Sapiens& 6&18& 8505& 137& 9& 12\\
587&Tomczak et al. &Mus musculus& 7&21& 860&50& 2& 20\\
586&Tomczak et al. &Mus musculus& 8&24& 5211& 118& 5& 20\\
\hline
\end{tabular*}   \vspace*{-3pt}
\end{table}

%

\begin{figure}

\includegraphics{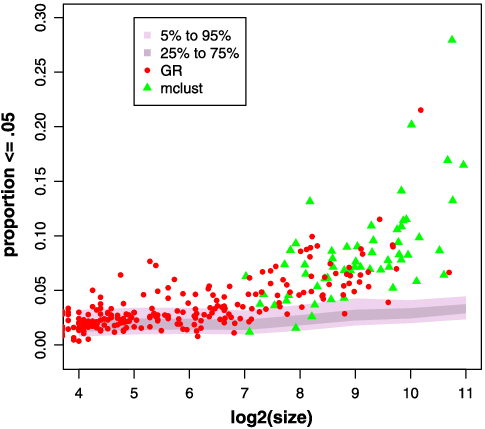}

\caption{Empirical study of the
association between clusters and biological
function. For every cluster identified by gamma ranking (red) or
\textup{mclust} (green) in
the data sets in Table~\protect\ref{tab:dat}, plotted is the proportion of small
enrichment p-values
(vertical) versus the cluster size (horizontal). The enrichment
p-values are
Fisher-exact-test p-values and the proportion is computed over a
database of GO and KEGG
pathways (Table \protect\ref{table7}). Bands indicate similar proportions computed for
random sets.}\label{figure4}
\end{figure}

The empirical study shows not only that gamma ranking produces substantially
different clusters than popular approaches, but also that the
identified clusters are significant in terms of their biological properties.
Investigators often measure the biological properties of a gene cluster by
identifying functional properties that seem to be over-represented in the
cluster. Gene set enrichment analysis is most frequently performed by
applying Fisher's exact test to each of a long list of functional categories,
testing the null hypothesis that the functional
category is independent of the gene cluster [e.g., \citet{newt07}].
Functional
categories from the Gene Ontology (GO) Consortium and the
Kyoto encyclopedia (KEGG) were used to assess
the biological properties of all the clusters
identified in the above calculation.
Specifically, we computed for each cluster a vector of $p$-values
across GO and KEGG. Figure~\ref{figure4}
shows the proportion of these $p$-values smaller
than 0.05, stratified by cluster size and in comparison to results on random
sets of the same size. Evidently, the clusters identified
by gamma ranking contain substantial biological information.

%

Figure~\ref{figure4} also shows that \texttt{mclust} clusters carry substantial biological
information, and a similar result is true for \texttt{K-means} (not shown).
Whatever cluster signal is present in the expression data, it is evident
that gamma-ranking finds different\vadjust{\goodbreak} aspects of this signal
than do the standard approaches,
while still delivering clusters that relate in some way to the biology.
Gamma-ranking clusters are not anonymous sets of genes with similar
expression profiles; they are sets of genes linked to an ordering pattern
in the underlying means. The commonly used clustering methods are
unsupervised, while gamma-ranking utilizes the known grouping labels
in the sample. It seems beneficial to use this grouping information;
undoubtedly various schemes could be developed. By their construction,
the gamma-ranking clusters have a simple interpretation in terms of
sets of genes supporting particular hypotheses about changes in
mean expression.

\section{Count data}\label{sec6}
Microarray technology naturally leads to continuous measurements of gene
expression, as modeled in Section~\ref{sec2}, but technological advances allow
investigators essentially to count the number of copies of each molecule
of interest in each sample [e.g., \citet{mo08}].
Poisson distributions are central in the analysis of such data [e.g.,
\citet{rna08}], and gamma ranking extends readily to this case.

Briefly, data at each gene (or tag) is a vector $x_g=(x_{g,1}, \ldots, x_{g,n})$
as before, but $x_{g,i}$ is now a count from the $i$th \textit{library}
(rather than an expression level on the $i$th microarray).
There may be replicate libraries within a given cellular state, and
comparisons of interest may be between different cellular states.
Library sizes~$\{N_i \}$, say, are additional
but known design parameters. Important parameters are expected counts
relative to some common library size.
Adopting the notation from Section~\ref{sec2}, a~cluster of libraries $\sigma(\eta,k)$ may share their
size-adjusted
expected values, and so for any $i \in\sigma(\eta, k)$ the observed count
$x_{g,i}$ arises from the Poisson distribution with mean $N_i \mu_k$.
Further, the structure $\eta$ on test puts an ordering constraint
$\mu_1 < \mu_2 <\cdots< \mu_{K_\eta} $ on these latent expectations.
The key is
to integrate away these latent expected values using a conjugate gamma
prior, but
conditionally on the ordering. Prior to conditioning, the $\mu_k$'s are
independent
and identical gamma variables with (integer) shape $\alpha_0$ and
rate $\alpha_0 \nu_0$. Then,
analogously to Theorem~\ref{theo2}, the predictive distribution
for the vector of
conditionally Poisson responses is
%
%
\begin{equation}
\label{eq:pg}
p(x_g|\eta) =
c_\eta \Biggl( \prod_{i=1}^n \frac{1}{ x_{g,i}!} \Biggr)
\underbrace{ \Biggl( \prod_{k=1}^{K_\eta} u_{g,k}
\Gamma( s_{g,k} + \alpha_0 ) \Biggr)
}_{ \mathrm{center} (\eta) }
P_{\mathrm{ord}}( \eta),
\end{equation}
where
\[
P_{\mathrm{ord}}(\eta) = P (Z_1 < Z_2 < \cdots< Z_{K_\eta} )
\]
with the $Z_k$'s mutually
independent gamma-distributed random variables with
(gene-specific) shapes $a_k = \alpha_0 + s_{g,k}$ and rates
$\lambda_k = \alpha_0 \nu_0 + n_{k}$. In~(\ref{eq:pg}), the
normalizing constant is
\[
c_\eta= \frac{K_\eta! (\alpha_0 \nu_0)^{ \alpha_0 K_\eta} }{
[ \Gamma(\alpha_0) ]^{ K_\eta} }
\prod_{k=1}^{K_\eta} ( \alpha_0 \nu_0 + n_{k} )^{-\alpha_0}
\]
and, further,
$s_{g,k} = \sum_{i \in\sigma(\eta,k) } x_{g,i}$, $n_{k} = \sum_{i \in
\sigma(\eta,k) }
N_i$ and
\[
u_{g,k} = \prod_{i \in\sigma(\eta,k) }
\biggl( \frac{ N_i }{ \alpha_0 \nu_0 + n_{k} } \biggr)^{x_{g,i}}.
\]
Notice that in $P_{\mathrm{ord}}(\eta)$ the event refers to an increasing
sequence of gamma's, rather
than a decreasing sequence, as in Theorem~\ref{theo1}. This arises because for
Poisson responses
the conjugate prior involves a gamma distribution for the means,
whereas for
gamma responses the conjugate is inverse gamma on the means. For
computations to work out,
the key thing is that some monotone transformation of each
latent mean has a gamma
distribution. In the null structure (all means equal),
$P_{\mathrm{ord}}
(\eta) = 1 $ and
(\ref{eq:pg}) reduces to the negative-multinomial distribution.
It will be important to study the practical
utility of~(\ref{eq:pg}) and overdispersed extensions [cf.
\citet{rs07}], but
such investigation
is not within the scope of the present paper. The main reason to
present the finding here is to show that gamma-rank probabilities
(Section~\ref{sec3}) arise
in multiple probability models.

\section{Concluding remarks}\label{sec7}

Calculations presented here consider a discrete mixture model and the resulting
clustering for gene-expression or similar data types. The discrete mixing
is over patterns of equality and inequality among latent expected values
(\textit{ordered structures}). Clustering by these patterns
addresses an important biological problem to organize gene expression
relative to various cellular states, which is part of the larger task
to determine biological function.
In examples the method was applied after a round of feature selection,
although it could have been applied to each full data set (i.e.,
by including
the null structure in the mix) and it could have been the basis of more
comprehensive analysis, going beyond clustering and towards hypothesis testing
and error-rate-controlled gene
lists. Our more conservative line is attributable in part to an
incomplete understanding of the method's robustness. Relaxing the
fixed-coefficient-of-variation assumption, as in \citet{logot06} or
\citet{ross09}, could be considered to address the problem. The focus
on clustering, however, is motivated largely by its practical utility
in the context of genomic data analysis.

By cataloging ordered structures, rather than the smaller set of
unordered structures, the mixture model produces readily interpretable
clusters in the multi-group setting.
\citet{jensen} argues similarly.
For example, the largest cluster of temporally responsive
genes in Edwards' data are upregulated immediately after
treatment and show no significant fluctuations thereafter.
The development of calculations for ordered structures has been more
challenging than for unordered structures, which were presented
in \citet{kend03} and implemented in the Bioconductor
package \texttt{EBarrays}.
Mixture calculations are simplified in the unordered
case because component densities reduce by factorization to
elementary products [i.e.,
the last factor in~(\ref{eq:gg}) is not present].
The requirement to compute gamma-rank probabilities
had limited a fuller development.

Gamma ranking produces clusters indexed by patterns of expected expression
rather than anonymous clusters defined by high similarity of
their contents. A~referee noted that large gamma-ranking clusters
may tend to swallow up genes more easily than small clusters because
the estimated posterior assignment probability is proportional to
the estimate of the mixing weight $\pi_\eta$: that is, structures with
large $\pi_\eta$ have a head start in the allocation of genes. On
one hand, this provides an efficiency which may be advantageous for
genes that have a relatively weak signal (and which otherwise
might be assigned to a more null-like structure). It also implies
that small clusters are more reliable, in a way, since the assigned
genes
have made it in spite of the small $\pi_\eta$. Another feature of
gamma ranking is that clusters can be tuned by a threshold parameter $c$,
as in~(\ref{eq:reg}), rather than being determined by Bayes's rule
assignment. Taking $c$ close to $1$ tends to purify the clusters;
the more equivocal genes drop into an unassigned category.
Empirically,
such swallowing up may not be substantial; at least in comparison
to the simpler clustering methods analyzed, gamma ranking produces more
and smaller clusters.

There is nothing explicit in gamma ranking that attends to the
temporal dependence which might seem to be involved in time-course data.
Independent cell lines were grown in the Edwards'
experiment, one for each microarray, and so there is independent
sampling in spite of a time component. Additionally, the model
imposes dependencies in~(\ref{eq:gg}) driven by whichever structure
$\eta$ governs data at a given gene. If there
were complicated temporal dependence, the identified clusters
would still reflect genes that act in concert in this experiment;
they might act in concert by a different $\eta$ in another run of the
experiment, and we would not be confident in the fitted proportions,
even though the clusters may continue to be informative.
Neither does the model have explicit dependence among genes;
but it produces clusters of genes that seem to be highly
associated (genes that realize the same structure $\eta$ seem to present
correlated data). This shows that a sufficiently rich hierarchical model,
based on lots of conditional independence, can represent
characteristics of dependent data. Of course, care is needed since
the sampling distribution of parameter estimates is affected
by the intrinsic dependencies within the data generating mechanism.

The mixture framework from \citet{kend03} has supported
a number of extensions to related problems in statistical genomics:
\citet{yuankend06b}
(time-course data), \citet{kend06} (mapping
expression traits) and \citet{keles07} (localizing transcription factors).
The ability to monitor ordered structures may have some application in these
problems. Further, the ability to compute gamma-rank probabilities may
have application in distinct inference problems [e.g., \citet{doksum}].
Future work includes developing a better software implementation
of gamma ranking, enabling the implementation to have additional
flexibility (e.g., gene-specific shapes $\alpha$), studying the method's
sampling properties and exploring extensions to emerging data sources.

\begin{appendix}
\section{\texorpdfstring{Proof of Theorem \lowercase{\protect\ref{theo2}}}{Proof of Theorem 2}}\label{appA}

Let $g_k(z)$ denote the density of a gamma
distribution with shape $a_k$ and rate $\lambda_k$. By definition
\begin{eqnarray*}
P(E)& =& \int_0^\infty\int_{z_{K}}^\infty
\cdots\int_{z_2}^\infty\Biggl[ \prod_{k=1}^{K} g_k(z_k)
\Biggr] \,dz_{1} \cdots dz_{K-1} \,dz_{K} \\
&=& \int_0^\infty g_{K}(z_{K}) \int_{z_{K}}^\infty
g_{K-1}(z_{K-1}) \cdots\int_{z_2}^\infty g_1(z_1) \,dz_1
\cdots dz_{K-1} \,dz_{K},
\end{eqnarray*}
where in the second line we move factors in the integrand
as far as possible to the left.
With this in mind we construct functions $f_k(z)$, $z \geq0$,
recursively as $f_0(z) = 1$ and, for $k=1,2, \ldots, K$,
%
%
\begin{equation}
\label{eq:recurse}
f_k(z) = \int_{z}^\infty f_{k-1}(u) g_{k}(u) \,du,
\end{equation}
and we observe that $P(E) = f_{K}(0)$. Evaluating these functions
further, we see
\begin{eqnarray*}
f_1(z) &=& \int_{z}^\infty g_1(z_1) \,dz_1 \\
&=& P ( Z_1 \geq z ) \\
&=& P ( M_1 < a_1 | Z_2 =z ) \\
&=& \sum_{m_1=0}^{a_1-1} {\mathrm{po}}(m_1; \lambda_1 z).
\end{eqnarray*}
Here $M_1 = \mathbb{M}_1(0,Z_2)$ is Poisson$(\lambda_1 z)$ distributed
conditionally upon $Z_2 = z$, and ${\mathrm{po}}(\cdot)$ indicates the
Poisson probability
mass function with the indicated parameter. The equivalence in the second
and third lines above stems from basic relationships between objects in the
underlying Poisson processes. As long as $M_1$ is small, it means that
the $\mathbb{N}_1$ process has not accumulated many points up to time
$Z_2=z$, and hence the $Z_1$ value must be relatively large. More
basically,
%
%
\begin{equation}
\label{eq:pgg}
P(U>u) = P(X<a),
\end{equation}
when $U \sim \operatorname{Gamma}(a, \lambda)$ and
$X \sim \operatorname{Poisson}(\lambda u)$, for integer shapes $a$.

Proceeding to $f_2(z)$,
\begin{eqnarray*}
f_2(z) &=& \int_{z}^\infty f_1(z_2) g_2(z_2) \,dz_2 \\
&=& \sum_{m_1=0}^{a_1-1} \int_z^\infty
\operatorname{po}(m_1; \lambda_1 z_2) g_2( z_2) \,dz_2 \\
&=& \sum_{m_1=0}^{a_1-1} p_1(m_1)
\int_{z}^\infty\frac{
\operatorname{po}(m_1; \lambda_1 z_2 ) ;, g_2(z_2)
}{ p_1(m_1) } \,dz_2.
\end{eqnarray*}
Here $p_1(m_1) $ is the probability mass function of a negative-binomial
distribution, as in~(\ref{eq:nb}). Indeed, we have reorganized the
summand above to highlight that integrand on the far right is precisely
the density function of a gamma distributed variable with
shape $a_2+m_1$ and rate $\lambda_1+\lambda_2$. This represents the
Poisson--Gamma conjugacy in ordinary Bayesian analysis [e.g.,
\citet{gelman04}, pages 52 and~53]. The
integral evaluates to $1$ if $z=0$, and hence we have proved the case\vadjust{\goodbreak}
$K=2$. But furthermore, the integral represents the chance that a
gamma distributed variable is large, and so by~(\ref{eq:pgg})
\begin{eqnarray*}
f_2(z) &=&
\sum_{m_1=0}^{a_1-1} \sum_{m_2=0}^{m_1+a_2-1} p_1(m_1)
\operatorname{po}\bigl(m_2; (\lambda_1+\lambda_2) z \bigr) \\
&=&
\sum_{m_1=0}^{a_1-1} \sum_{m_2=0}^{m_1+a_2-1} p_1(m_1)
\operatorname{po}(m_2; \Lambda_2 z ).
\end{eqnarray*}
The baseline result of an induction proof has been established.
Assume that for some $k \geq3$,
%
%
\begin{equation}
\label{eq:induct}
f_{k-1}(z) = \sum_{m_1=0}^{a_1-1} \cdots\sum_{m_{k-1}=0}^{m_{k-2}+a_{k-1}-1}
\Biggl( \prod_{j=1}^{k-2} p_j(m_j) \Biggr)
\operatorname{po}(m_{k-1}; \Lambda_{k-1} z )
\end{equation}
and then evaluate~(\ref{eq:recurse}) to obtain
\begin{eqnarray*}
f_k(z) &=& \int_z^\infty f_{k-1}(z_k) g_k(z_k) \,dz_k \\
&=& \int_z^\infty
\sum_{m_1=0}^{a_1-1} \cdots\sum_{m_{k-1}=0}^{m_{k-2}+a_{k-1}-1}
\Biggl( \prod_{j=1}^{k-2} p_j(m_j) \Biggr)
\operatorname{po}(m_{k-1}; \Lambda_{k-1} z_k ) g_k(z_k) \,dz_k \\
&=&
\sum_{m_1=0}^{a_1-1} \cdots\sum_{m_{k-1}=0}^{m_{k-2}+a_{k-1}-1}
\Biggl( \prod_{j=1}^{k-2} p_j(m_j) \Biggr)
\int_z^\infty\operatorname{po}(m_{k-1}; \Lambda_{k-1} z_k ) g_k(z_k)\,
dz_k \\
&=&
\sum_{m_1=0}^{a_1-1} \cdots\sum_{m_{k-1}=0}^{m_{k-2}+a_{k-1}-1}
\Biggl( \prod_{j=1}^{k-1} p_j(m_j) \Biggr)
\int_z^\infty
\frac{ \operatorname{po}(m_{k-1}; \Lambda_{k-1} z_k ) g_k(z_k) }{
p_{k-1}(m_{k-1})}\,
dz_k \\
&=&
\sum_{m_1=0}^{a_1-1} \cdots\sum_{m_{k-1}=0}^{m_{k-2}+a_{k-1}-1}
\Biggl( \prod_{j=1}^{k-1} p_j(m_j) \Biggr)
\sum_{m_k=0}^{m_{k-1}+a_k-1} \operatorname{po}( m_k; \Lambda_k z ) \\
&=&
\sum_{m_1=0}^{a_1-1} \cdots\sum_{m_k=0}^{m_{k-1}+a_k-1}
\Biggl( \prod_{j=1}^{k-1} p_j(m_j) \Biggr)
\operatorname{po}( m_k; \Lambda_k z ),
\end{eqnarray*}
which establishes that~(\ref{eq:induct}) is true for \textit{all} $k$.
Evaluating at $k=K$ and $z=0$ establishes the theorem.

\textit{Coda}: Further insight is gained by realizing from the definition
of the
counts that
\begin{eqnarray*}
\mathbb{M}_k(Z_k) &=& \mathbb{M}_{k-1}(Z_k) + a_k \\
&=& M_{k-1} + a_k.
\end{eqnarray*}
But also $\mathbb{M}_k$ has a jump at $Z_k$, and so we see
the equivalence
%
%
\begin{equation}
Z_{k} > Z_{k+1} \quad\Longleftrightarrow\quad M_k < M_{k-1} + a_k.
\end{equation}
The event $E$ is an intersection of these pairwise events, and this is
manifested in the ranges of summation in~(\ref{eq:f1}). In contrast
to~(\ref{eq:f1}), these event considerations give $P(E)$ equal to
%
%
\begin{equation}
\label{eq:f3}
\sum_{m_1=0}^{a_1 - 1} \sum_{m_2=0}^{m_1+a_2 -1 }
\cdots\sum_{m_{K-1} = 0 }^{m_{K-2} + a_{K-1} - 1 }
p_{\mathrm{joint}}(m_1,m_2, \ldots, m_{K-1}).
\end{equation}
The implication seems to be
that $M_1, M_2, \ldots, M_{K-1}$ are mutually independent,
though Theorem~\ref{theo1} does not confirm this because the factorization into
negative binomials
is required for all arguments, beyond what is shown.
It is a conjecture that the $\{ M_k \}$ are
mutually independent. A~proof by brute force evaluation
in the special cases $K=3$ and
$K=4$ is available (not shown),
but we have not found a general proof. The fact is somewhat surprising
because the $\{ \mathbb{M}_k \}$ processes
are highly positively dependent. The
independence seems to emerge as a balance between this positive dependence
and the negative association created by $Z_k$ being inversely related to
$\mathbb{M}_k(0,t]$.

\section{\texorpdfstring{Linear independence and~proof~of~Theorem~\lowercase{\protect\ref{theo3}}}
{Linear independence and proof of Theorem 3}}\label{appB}

Consider the three-dimensional case, and initially consider a single
replicate in each of the three groups.
Data on each gene form the vector $(x,y,z)$, say, of three positive reals.
Thirteen component densities $p(x,y,z|\eta)$ constitute the mixture
model (Table~\ref{tab:pg10}).
For a vector $a=(a_\eta)$ of reals, the test function is
$ T_a(x,y,z) = \sum_\eta a_\eta p(x,y,z|\eta) $.
It needs to be shown that if $T_a(x,y,z) = 0$ for all $x,y,z>0$, then
$a_\eta=0$ for
all structures $\eta$. Specializing~(\ref{eq:gg}) to this case,
and eliminating the positive factor $(xyz)^{\alpha-1}$, we see
that $T_a(x,y,z)=0$ is equivalent to
%
%
\begin{equation}
\label{eq:d1}
\sum_{\eta} a_\eta c_\eta \operatorname{center}(\eta) P_{\mathrm
{ord}}(\eta) = 0.
\end{equation}

%
\begin{table}[h]
\caption{Thirteen structured components $p(x,y,z|\eta) = c_\eta
(xyz)^{\alpha-1}
\operatorname{center}(\eta) P_{\mathrm{ord}}(\eta)$ in the three dimensional,
no-replicate case. The forms have been simplified, w.l.o.g., by taking the~scale $\nu_0=1$, by
writing $\beta= \alpha_0+\alpha$ and $\xi= \alpha_0/\alpha$. Normalizing
constants $c_\eta$ are~as~in~(\protect\ref{eq:gg}). Note that the
$e_m$ and $e_{m,n}$ stand for constants (not involving $x,y,z$), 
but~possibly~differing among rows}\label{tab:pg10}
\begin{tabular*}{\tablewidth}{@{\extracolsep{\fill}}lcc@{}}
\hline
\multicolumn{1}{@{}l}{\textbf{Structure} $\bolds\eta$}
& $\bolds{[ \operatorname{center}(\eta
) ]^{-1}}$
& $\bolds{P_{\mathrm{ord}}(\eta)}$ \\
\hline
$(123)$ & $(x+y+z+\xi)^{\beta+ 2 \alpha} $ & $1$ \\
[2pt]
$(12)(3)$ & $(x+y+\xi)^{\beta+\alpha} (z+\xi)^{\beta}$ &
$ \sum_{m=0}^{\beta+ \alpha-1} \frac{ e_m ( z + \xi)^{\beta}
(x+y+\xi)^{m} }{ (x+y+z+2\xi)^{\beta+m }} $ \\
[6pt]
$ (3)(12)$ & `` &
$ \sum_{m=0}^{\beta-1} \frac{ e_m ( z + \xi)^{m}
(x+y+\xi)^{\beta+\alpha} }{ (x+y+z+2\xi)^{\beta+ \alpha+m }} $ \\
[6pt]
$(13)(2)$ & $(x+z+\xi)^{\beta+\alpha} (y+\xi)^{\beta}$ &
$ \sum_{m=0}^{\beta+ \alpha-1} \frac{ e_m ( y + \xi)^{\beta}
(x+z+\xi)^{m} }{ (x+y+z+2\xi)^{\beta+m }} $ \\
[6pt]
$ (2)(13)$ & `` &
$ \sum_{m=0}^{\beta-1} \frac{ e_m ( y + \xi)^{m}
(x+z+\xi)^{\beta+\alpha} }{ (x+y+z+2\xi)^{\beta+ \alpha+m }} $ \\
[6pt]
$(1)(23)$ & $(y+z+\xi)^{\beta+\alpha} (x+\xi)^{\beta}$ &
$ \sum_{m=0}^{\beta-1} \frac{ e_m ( x + \xi)^{m}
(y+z+\xi)^{\beta+\alpha} }{ (x+y+z+2\xi)^{\beta+ \alpha+m }} $ \\
[6pt]
$(23)(1)$ & `` &
$ \sum_{m=0}^{\beta+ \alpha-1} \frac{ e_m ( x + \xi)^{\beta}
(y+z+\xi)^{m} }{ (x+y+z+2\xi)^{\beta+m }} $ \\
[6pt]
$(1)(2)(3)$ & $[ (x+\xi)(y+\xi)(z+\xi) ]^{\beta}$ &
$ \sum_{m=0}^{\beta-1} \sum_{n=0}^{m+\beta-1}
\frac{e_{m,n} (x+\xi)^m [ (y+\xi)(z+\xi)] ^{ \beta}
(x+y+2\xi)^{n} }{
(x+y+2\xi)^{\beta+m} (x+y+z+3\xi)^{\beta+n} } $ \\
[6pt]
$(2)(1)(3)$ & `` &
$ \sum_{m=0}^{\beta-1} \sum_{n=0}^{m+\beta-1}
\frac{e_{m,n} (y+\xi)^m [ (x+\xi)(z+\xi)] ^{ \beta}
(x+y+2\xi)^{n} }{
(x+y+2\xi)^{\beta+m} (x+y+z+3\xi)^{\beta+n} } $ \\
[6pt]
$(1)(3)(2)$ & `` &
$ \sum_{m=0}^{\beta-1} \sum_{n=0}^{m+\beta-1}
\frac{e_{m,n} (x+\xi)^m [ (y+\xi)(z+\xi)] ^{ \beta}
(x+z+2\xi)^{n} }{
(x+z+2\xi)^{\beta+m} (x+y+z+3\xi)^{\beta+n} } $ \\
[6pt]
$(2)(3)(1)$ & `` &
$ \sum_{m=0}^{\beta-1} \sum_{n=0}^{m+\beta-1}
\frac{e_{m,n} (y+\xi)^m [ (z+\xi)(x+\xi)] ^{ \beta}
(y+z+2\xi)^{n} }{
(y+z+2\xi)^{\beta+m} (x+y+z+3\xi)^{\beta+n} } $ \\
[6pt]
$(3)(1)(2)$ & `` &
$ \sum_{m=0}^{\beta-1} \sum_{n=0}^{m+\beta-1}
\frac{e_{m,n} (z+\xi)^m [ (x+\xi)(y+\xi)] ^{ \beta}
(x+z+2\xi)^{n} }{
(x+z+2\xi)^{\beta+m} (x+y+z+3\xi)^{\beta+n} } $ \\
[6pt]
$(3)(2)(1)$ & `` &
$ \sum_{m=0}^{\beta-1} \sum_{n=0}^{m+\beta-1}
\frac{e_{m,n} (z+\xi)^m [ (x+\xi)(y+\xi)] ^{ \beta}
(y+z+2\xi)^{n} }{
(y+z+2\xi)^{\beta+m} (x+y+z+3\xi)^{\beta+n} } $ \\
\hline
\end{tabular*}
\end{table}

A strictly positive multivariate polynomial $\phi(x,y,z)$ is required
that can convert the left-hand side of~(\ref{eq:d1}) into a polynomial
by the canceling of denominator factors. Specifically, $\phi= \phi_1
\phi_2$
where $\phi_1(x,y,z)$ controls factors in $\operatorname{center}(\eta)$ and
$\phi_2(x,y,z)$ controls factors in $P_{\mathrm{ord}}(\eta)$.
Inspection suggests
taking $\phi_1(x,y,z)$ equal to
\[
(x+y+z+\xi)^{\beta+2\alpha} [ (x+y+\xi)
(x+z+\xi)(y+z+\xi) ]^{\beta+ \alpha}
[ (x+\xi)(y+\xi)(z+\xi) ]^{\beta}
\]
and $\phi_2(x,y,z)$ equal to
\begin{eqnarray*}
&& (x+y+z + 2\xi)^{2\beta+\alpha-1}
[ (x+y+2\xi)(x+z+2\xi)(y+z+2\xi) ]^{2\beta-1 } \\
&&\qquad{} \times
( x+y+z+3\xi)^{3\beta-2}.
\end{eqnarray*}
Observe that the degree of $x$ in the polynomial $\phi=\phi_1 \phi_2$
is \mbox{$13\beta+5\alpha-5$}. Indeed this is also the degree of $y$ and the
degree of $z$ by symmetry. These degrees are reduced in the polynomial
$f_\eta= \phi(x,y,z)
\operatorname{center}( \eta) P_{\mathrm{ord}}(\eta)$ by factors
in the denominators of $\operatorname{center}(\eta)$ and $P_{\mathrm
{ord}}(\eta)$.
For example, if $\eta=(12)(3)$, then
\begin{eqnarray*}
f_\eta&=&
(x+y+z+\xi)^{\beta+2\alpha} [
(x+z+\xi)(y+z+\xi) ]^{\beta+ \alpha}
[ (x+\xi)(y+\xi) ]^{\beta} \\
&&{}\times
[ (x+y+2\xi)(x+z+2\xi)(y+z+2\xi) ]^{2\beta-1 }
( x+y+z+3\xi)^{3\beta-2} \\
&&{} \times
\sum_{m=0}^{\beta+ \alpha-1} e_m ( z + \xi)^{\beta}
(x+y+\xi)^{m} (x+y+z+2\xi)^{\beta+\alpha-1 - m },
\end{eqnarray*}
which is a polynomial of
degree $11\beta+4\alpha-5$, in both $x$ and $y$, and
of degree $12 \beta+ 5\alpha-5$ in $z$. A~similar construction is possible
for all structures; Table~\ref{tab3} records the degrees
of $x$, $y$ and $z$ in all component polynomials $f_\eta$.

%
\begin{table}
\caption{Degree of $x,y$ and $z$ in the multivariate polynomials
$f_\eta= \phi(x,y,z)
\operatorname{center}(\eta) P_{\mathrm{ord}}(\eta)$. Recall $\beta
=\alpha_0 +
\alpha$ and both $\alpha$ and $\alpha_0$ are positive integers} \label{tab3}
\begin{tabular*}{\tablewidth}{@{\extracolsep{\fill}}lccc@{}}
\hline
\textbf{Structure} $\bolds{\eta}$ & \textbf{Degree}$\bolds{(x)}$
& \textbf{Degree}$\bolds{(y)}$ & \textbf{Degree}$\bolds{(z)}$ \\
\hline
$(123)$ & $11\beta+4\alpha-4$ & $11\beta+4\alpha-4$ & $11\beta+4\alpha
-4$\\
$(12)(3)$ & $11\beta+4\alpha-5$ & $11\beta+4\alpha-5$ & $12\beta
+5\alpha-5$\\
$ (3)(12)$& $12\beta+4\alpha-5$ & $12\beta+4\alpha-5$ & $11\beta
+4\alpha-5$\\
$(13)(2)$ & $11\beta+4\alpha-5$ & $12\beta+5\alpha-5$ & $11\beta
+4\alpha-5$\\
$ (2)(13)$ &$12\beta+4\alpha-5$ & $11\beta+4\alpha-5$ & $12\beta
+4\alpha-5$\\
$(1)(23)$ & $11\beta+4\alpha-5$ & $12\beta+4\alpha-5$ & $12\beta
+4\alpha-5$\\
$(23)(1)$ & $12\beta+5\alpha-5$ & $11\beta+4\alpha-5$ & $11\beta
+4\alpha-5$\\
$(1)(2)(3)$&$10\beta+5\alpha-5$ & $11\beta+5\alpha-5$ & $12\beta
+5\alpha-5$\\
$(2)(1)(3)$&$11\beta+5\alpha-5$ & $10\beta+5\alpha-5$ & $12\beta
+5\alpha-5$\\
$(1)(3)(2)$&$10\beta+5\alpha-5$ & $12\beta+5\alpha-5$ & $11\beta
+5\alpha-5$\\
$(2)(3)(1)$&$12\beta+5\alpha-5$ & $10\beta+5\alpha-5$ & $11\beta
+5\alpha-5$\\
$(3)(1)(2)$&$11\beta+5\alpha-5$ & $12\beta+5\alpha-5$ & $10\beta
+5\alpha-5$\\
$(3)(2)(1)$&$12\beta+5\alpha-5$ & $11\beta+5\alpha-5$ & $10\beta
+5\alpha-5$\\
\hline
\end{tabular*}
\end{table}

Having introduced the multiplier $\phi$, the linear independence~(\ref{eq:d1})
is equivalent to the assertion that polynomial equation
%
%
\begin{equation}
\label{eq:d2}
\sum_\eta a_\eta c_\eta f_\eta(x,y,z) = 0\qquad
\mbox{for all $x,y,z >0$}\vadjust{\goodbreak}
\end{equation}
implies $a_\eta=0$ for all $\eta$. Fixing any $y$ and $z$, the
left-hand side of
equation~(\ref{eq:d2}) is a polynomial in $x$, with degree $12\beta+
5 \alpha-5$, according
to Table~\ref{tab3}. Indeed terms associated with structures
$\eta= (23)(1)$, $(2)(3)(1)$ and $(3)(2)(1)$ all contribute monomials
with that highest power in $x$. The coefficient of $x^{12\beta+5\alpha-5}$,
denoted $d=d(a,y,z)$, equals
\[
a_{(23)(1)} c_{(23)(1) } f'_{(23)(1)} +
a_{(2)(3)(1)} c_{(2)(3)(1) } f'_{(2)(3)(1)} +
a_{(3)(2)(1)} c_{(3)(2)(1) } f'_{(2)(3)(1)},
\]
where $f'$ indicates contributions from respective terms within $f_\eta$.
This coefficient $d$ must equal zero, for all $y$ and $z$;
after all, a~degree $12\beta+ 5 \alpha-5$ polynomial
can equal zero in $x$ for at most
that many $x$ values, unless the coefficient $d$ is exactly zero; and
we are asking
that it equal zero at all values of $x$. From this study of the high-power
coefficient in $x$, we have reduced consideration
to three structures and are able to focus on
$d=d(a,y,z)$ as a bivariate polynomial in $y$ and $z$ (Table~\ref{tab4}).

%
\begin{table}
\tablewidth=250pt
\caption{Degrees of $y$ and $z$ in three terms of the bivariate polynomial
$d(a,y,z)$. This is a subset of Table \protect\ref{tab3}}\label{tab4}
\begin{tabular*}{\tablewidth}{@{\extracolsep{\fill}}lcc@{}}
\hline
\textbf{Structure} $\bolds{\eta}$ & \textbf{Degree}$\bolds{(y)}$
& \textbf{Degree}$\bolds{(z)}$ \\
\hline
$(23)(1)$ & $11\beta+4\alpha-5$ & $11\beta+4\alpha-5$\\
$(2)(3)(1)$& $10\beta+5\alpha-5$ & $11\beta+5\alpha-5$\\
$(3)(2)(1)$& $11\beta+5\alpha-5$ & $10\beta+5\alpha-5$\\
\hline
\end{tabular*}
\end{table}

The initial argument focusing on the degree of $x$ can
be adapted to study other variables in Table~\ref{tab4}.
With degree of $y$ equal to $11\beta+ 5\alpha-5$, for instance,
it must be that the coefficient
$d'(z)$, say, of
$y^{11\beta+5\alpha-5}$ equals zero for all $z$. After all, the
polynomial can
equal zero at at most $11\beta+5\alpha-5$ $y$ values, and we require
it to be zero at all $y$. But all contributions
to that coefficient are strictly positive, except possibly $a_{(3)(2)(1)}$,
hence we conclude $a_{(3)(2)(1)} = 0$. By the same token, working with
the degree $11\beta+5\alpha-5$ term in $z$, it follows that
$a_{(2)(3)(1)} = 0$,
which then forces $a_{(23)(1)}=0$, because we require $d=0$ overall.
Three rows from Table~\ref{tab3} have been eliminated (i.e., forced
$a_\eta=0$), all those in which
the mean of the first variable is
greater than the other two means. Next, return to
the reduced table, and focus, say, on structures $(13)(2)$, $(1)(3)(2)$
and $(3)(1)(2)$, in which the second variable has mean greater than
the others.
In doing so, three more coefficients $a_{(13)(2)} = a_{(1)(3)(2)}
= a_{(3)(2)(1)} = 0$ are forced, and Table~\ref{tab:dat} is further reduced
to seven rows. Then
the argument is repeated to get \mbox{$a_{(12)(3)} = a_{(1)(2)(3)} =
a_{(2)(1)(3)} = 0$},
and it remains to assess coefficients $a_\eta$ of
the four structures in Table~\ref{tab5}.

%
\begin{table}[b]
\tablewidth=250pt
\caption{Final subtable}\label{tab5}
\begin{tabular*}{\tablewidth}{@{\extracolsep{\fill}}lccc@{}}
\hline
\textbf{Structure} $\bolds{\eta}$ & \textbf{Degree}$\bolds{(x)}$ &
\textbf{Degree}$\bolds{(y)}$ & \textbf{Degree}$\bolds{(z)}$ \\
\hline
$(123)$ & $11\beta+4\alpha-4$ & $11\beta+4\alpha-4$ & $11\beta+4\alpha
-4$\\
$ (3)(12)$& $12\beta+4\alpha-5$ & $12\beta+4\alpha-5$ & $11\beta
+4\alpha-5$\\
$ (2)(13)$ &$12\beta+4\alpha-5$ & $11\beta+4\alpha-5$ & $12\beta
+4\alpha-5$\\
$(1)(23)$ & $11\beta+4\alpha-5$ & $12\beta+4\alpha-5$ & $12\beta
+4\alpha-5$\\
\hline
\end{tabular*}
\end{table}

The argument is repeated in this domain, knowing that all but four terms
in~(\ref{eq:d2}) have been eliminated.
The degree of $x$ is $12\beta+4\alpha-5$,
and there are contributions from both $\eta=(3)(12)$ and $\eta=(2)(13)$.
But then
restricted to these rows we get $a_{(3)(12)}=0$ because the
coefficient of $x^{12\beta+4\alpha-5}$ as a polynomial in $y$ has
degree $12\beta+4 \alpha-5$. The remaining constants $a_\eta$ are
similarly zero, completing the proof in the no-replicate ($m=1$),
three group $(p=3)$ case.

The balanced three group case follows suit, noting that
now $x$, $y$ and $z$ are sums taken, respectively, across
replicates in each of the three groups. The product statistic is
not $xyz$, but anyway it is common to all components and thus cancels
in the linear combination test function. The observation-related
shape parameter $\alpha$ is replaced by $m\alpha$. The two-dimensional
$(p=2)$ case is simpler and is left as an exercise.\

\section{\texorpdfstring{Strict concavity of log-likelihood and proof of Theorem \lowercase{\protect\ref{theo4}}}
{Strict concavity of log-likelihood and proof of Theorem 4}}\label{appC}

Let $q$ denote the number of nonnull structures, and
consider the log-likelihood $l(\pi)$ in~(\ref{eq:loglik}) to be
on $\mathbb{R}^q$, with the null probability defined
secondarily as $\pi_{0} = 1 - \sum_{\eta\neq\eta_0} \pi_\eta$.
This way we need not invoke Lagrange multipliers to compute derivatives of
$l(\pi)$. By calculus, the $q\times q$ Hessian $H$ of negative 2$nd$ derivatives
of $l(\pi)$ has $(ij)$th entry
\begin{eqnarray*}
H_{ij} &=& \sum_g \frac{ [ p(x_g|\eta_i) - p(x_g|\eta_0) ]
[ p(x_g|\eta_j) - p(x_g| \eta_0) ] }
{[ p(x_g)]^2} \\
&=& \sum_g f_i( x_g ) f_j(x_g),
\end{eqnarray*}
where $p(x_g)$ is the marginal density obtained by mixing over structures,
as in~(\ref{eq:mix}), and $f_i(x) = [ p(x|\eta_i) - p(x|\eta_0)
]/
p(x) $. Now let $a=(a_\eta)$ be a $q$-vector of constants. To determine
curvature of the log-likelihood we consider the quadratic form
\begin{eqnarray*}
a^T H a &=& \sum_{i=1}^q \sum_{j=1}^q
a_i a_j \sum_g f_i(x_g) f_j(x_g)
= \sum_g \Biggl( \sum_{i=1}^q a_i f_i(x_g) \Biggr)^2 \\
&=& \sum_g [ T_a(x_g) ]^2,
\end{eqnarray*}
where $T_a(x) = \sum_{i=1}^q a_i f_i(x)$. Clearly, $a^T H a \geq0$ regardless
of $a$ and so $H$ is nonnegative definite and $l(\pi)$ is concave.
To establish \textit{strict} concavity requires that we show $T_a(x_g) =
0$ for
all $g$ if and only if $a=0$. The following lemma shows that knowing
$T_a(x_g)=0$
for all $G$ values $x_g$ is enough to force $T_a(x)=0$ for all~$x$, as long
as $G$ is sufficiently large. But then $a=0$ by the linear independence
assumption, completing the proof.
\begin{lemma}\label{lemma1}
Let $\psi(x)$ be a multivariate polynomial in $x \in\mathbb{R}^n$, and
let $X_1, X_2, \ldots, X_m$ denote a random sample from a continuous
distribution on $\mathbb{R}^n$. If $m$ is at least as large as the number
of monomials in $\psi$, then, with probability one, $\psi(X_i) = 0$
for $i=1,2,
\ldots, m$ implies $\psi(x) = 0$ for all $x$.
\end{lemma}
\begin{pf}
Every point $X_i$ puts a linear condition on
the space
of coefficients of~$\psi$. It needs to be verified that these
conditions are
linearly independent. Suppose that the first $k$ conditions are linearly
independent, so the space of $\psi$'s that are zero at $X_1, \ldots, X_k$
has dimension (number of monomials in $\psi$) minus $k$. Pick one such
nonzero polynomial and call it $\phi$. Since $\phi= 0$ is a set
with positive codimension, we may assume (with probability one)
that
$\phi(X_{k+1})$
is not zero. Then, if we impose the additional condition $\psi(X_{k+1})=0$,
the dimension of the solution space drops by at least one, hence it drops
by one. Letting $k$ increase from $1$ to $m$ completes the proof.~%
\end{pf}

%
\begin{table}[b]
\tablewidth=300pt
\caption{Parameter estimates (not including mixing proportions) from
the examples analyzed.
The last column indicates the number of functional categories
in GO and KEGG having at least five annotated genes, which were used
in the development of Figure \protect\ref{figure4}. KEGG was not available for GDS1937,
and so this data set was not used in Figure \protect\ref{figure4}}\label{table7}
\begin{tabular*}{\tablewidth}{@{\extracolsep{\fill}}ld{3.0}cd{4.2}c@{}}
\hline
\textbf{Data set} & \multicolumn{1}{c}{$\bolds{\alpha}$}
& \multicolumn{1}{c}{$\bolds{\alpha_0}$} & \multicolumn{1}{c}{$\bolds{\nu_0}$}
& \textbf{\# GO/KEGG}
\\
\hline
Edwards & 113 & 1 & 586.5 & \\
GDS2323 & 14 & 1 & 119.1& 3849/184 \\
GDS1802 & 17 & 1 & 46.8& 3619/182 \\
GDS2043 & 22 & 1 & 47.7& 3619/182 \\
GDS2360 & 8 & 1 & 15.8& 3258/175 \\
GDS599 & 12 & 1 & 0.01& 3180/159 \\
GDS812 & 5 & 1 & 15.4& 3258/175 \\
GDS1937 & 6 & 1 & 20.5& NA \\
GDS568 & 10 & 1 & 37.1& 3258/175 \\
GDS2431 & 4 & 1 & 67.6& 4085/188 \\
GDS587 & 8 & 1 & 9999.2 & 1876/127 \\
GDS586 & 13 & 1 & 4566.2& 3258/175 \\
\hline
\end{tabular*}
\end{table}

\section{Further details of numerical examples}\label{appD}

The parameters $\alpha, \alpha_0$ and $\nu_0$
were fixed at values obtained by first fitting the unordered gamma--gamma
mixture model in \texttt{EBarrays}, without a null structure but
otherwise allowing all possible unordered structures.
Shape parameters were then rounded
to the nearest positive integer (Table~\ref{table7}) and all three
parameters were plugged into the EM procedure to
fit the proposed mixture-model proportions. [Recall that $P_{\mathrm
{ord}}(\eta)$
in~(\ref{eq:gg}) can be computed only for integer shapes, hence the
rounding.]
To simplify EM calculations in the four examples having more than
five groups, the full set of ordered structures was filtered to
a reduced set based on the fitting of the unordered gamma--gamma model
in \texttt{EBarrays}. Each ordered structure corresponds to exactly
one unordered structure (a many to one mapping).
If no gene had a high (greater than 0.5)
probability of mapping to a given unordered structure, then we deemed all
corresponding ordered structures to have $\pi_\eta= 0$. This approximation
is not ideal, since the Bayes rule assignment for some genes may
be one of the structures eliminated by forcing $\pi_\eta=0$. This
affects only 29 genes out of the 17\mbox{,}539 clustered in these four cases.
It would not affect clustering by a high threshold.

For all data sets, we examined quantile--quantile
plots and plots relating sample coefficient of variation to sample mean.
Some model violations were noted, but largely the gamma observation
model was supported.

For the Edwards data, we reran the EM algorithm for 10 cycles
and updated shape parameter
estimates via 2D grid search in each cycle. Estimated shapes changed
slightly; 784/786 genes received the same Bayes rule cluster
assignment.

Computations were done in R on industry-standard linux machines. For the
data sets analyzed, run times ranged from 6 to 860 CPU seconds per EM
iteration, with a mean of 270 seconds. Run time is affected by the number
of genes analyzed, the number of groups and also the shape parameters
and sample sizes.
\end{appendix}

\section*{Note added in proof}
The authors acknowledge
that the finding in Theorem~\ref{theo2} was reported previously by R. J. Henery (1983)
\textit{Journal of Applied Probability} \textbf{20} 822--834.

\section*{Acknowledgments}
We thank Lev Borisov for the proof of Lemma~\ref{lemma1},
Christina Kendziorski, an Associate Editor and two referees
for comments that greatly improved the development
of this work.

%
\printaddresses

\end{document}